\documentclass[journal]{IEEEtran}

\usepackage[dvips]{graphicx}
\usepackage{amsmath,amssymb,amsfonts}
\usepackage{amsthm}
\usepackage{epsf,epsfig}
\usepackage{color}
\usepackage{cite}
\usepackage{stfloats}
\usepackage[caption=false,font=footnotesize]{subfig}
\newtheorem{theorem}{Theorem}

\newtheorem{lemma}{Lemma}

\begin{document}
\title{ Cram\'er-Rao Lower Bound for DoA Estimation with RF Lens-Embedded Antenna Array}

\author{\IEEEauthorblockN{Jae-Nam~Shim,~Hongseok~Park,~GeeYong Suk,~\IEEEmembership{Student Member,~IEEE,} \\Chan-Byoung~Chae,~\IEEEmembership{Senior Member,~IEEE,}~and Dong Ku Kim,~\IEEEmembership{Senior Member,~IEEE}}
\thanks{J.-N Shim, H. Park, D. Kim are with the School of Electrical and Electronic Engineering, Yonsei University, Seoul 03722, Korea. (e-mail: \{jaenam, phs0127, dkkim\} @yonsei.ac.kr).}
\thanks{G. Suk and C.-B. Chae is with the with the School of Integrated Technology, Yonsei University, Seoul 03722, Korea (e-mail: renightmare2@naver.com, \{gysuk, cbchae\}@yonsei.ac.kr).}
}

\markboth{Draft Version}
{Shell \MakeLowercase{\textit{et al.}}}

\maketitle
\IEEEpeerreviewmaketitle

\begin{abstract}
In this paper, we consider the Cram\'er-Rao lower bound (CRLB) for estimation of a lens-embedded antenna array with deterministic parameters. Unlike CRLB of uniform linear array (ULA), it is noted that CRLB for direction of arrival (DoA) of lens-embedded antenna array is dominated by not only angle but characteristics of lens. Derivation is based on the approximation that amplitude of received signal with lens is approximated to Gaussian function. We confirmed that parameters needed to design a lens can be derived by standard deviation of Gaussian, which represents characteristic of received signal, by simulation of beam propagation method. Well-designed lens antenna shows better performance than ULA in terms of estimating DoA. This is a useful derivation because, result can be the guideline for designing parameters of lens to satisfy certain purpose.

\end{abstract}


\section{Introduction}

Reflecting the advent of a 5th generation (5G) communication, key technologies for next generation communication is under discussion. Extremely high data rate and low latency are some of the most notable changes in communication environment, to resolve those, massive multiple-input multiple-output (MIMO), millimeter-wave and densification are mentioned as core technologies in \cite{Andrews:2014ez}, \cite{Boccardi:2014em}. Additionally, a novel approach to apply lens antenna to mmWave MIMO, which is commonly used for radar and satellite communication systems, was proposed in \cite{Zeng:2014je} based on advantages of high gain, narrow beamwidth and low sidelines in different directions. 

In \cite{Kela:2016td}, it is proved that in ultra dense networks, beamforming utilizing location information based on line-of-sight (LOS) is superior compared to beamforming based on full band channel state information in terms of user throughput regardless of mobility of users, where the angles of departure and arrival is assumed to be given with respect to local coordinate systems, however, angles are obtained by acquisition of direction and distance information which is estimated by direction of departure (DoD), direction of arrival (DoA), time of arrival (ToA) and time of departure (ToD) in practice. The location information is one of the core elements for 5G communication in throughput boosting perspective. 

With these overall superiorities of using DoA based beamforming, we focus on DoA estimation which constructs the location information exploiting characteristic of received signal of lens antenna, whereas existing papers concerned about transmission and reception. In this paper, we compare error bound of DoA estimation  lens antenna based on preceding research in  \cite{Kwon:2016dz} for uniform linear array (ULA). Cram\'er Rao lower bound (CRLB), well known bound on the variance of estimator of a deterministic parameter valid for unbiased estimator with additive Gaussian noise assumption, was used for criterion of comparison.

\section{Lens Characteristic} \label{sec:lens model}
The characteristic of lens is defined by four parameters, $f$, $D$, $T$ and $\epsilon_{r}$, which the focal length, the aperture diameter, thickness and electric permittivity of lens respectively. Since our goal is to find the lower bound for error variance of DoA estimation, considering all of parameters is unnecessary and burden. All we need to derive the bound is the received signal of RF lens embedded antenna with expression of deterministic parameters.

 The array response of lens antenna is determined by characteristic of lens and distance between lens and antenna. It was modeled as sinc function \cite{Zeng:2016jl}, where antenna elements are placed along the arc whose diameter is the same as the focal length. With lens assisted linear array, it is experimentally measured in \cite{Kwon:2016dz}, that the estimated coefficient of lens antenna response ${\bf{a}}(y;\phi)$ can be modeled as 1-Gaussian fitting model and each parameter of the fitted gaussian, $p(\phi)$, $q(\phi)$ and $r(\phi)$, are provided.  
\begin{equation}\label{amplitude}
{\bf{a}}(y;\phi)=p(\phi)e^{-(\frac{y-q(\phi)}{r(\phi)})^2}
\end{equation}
 We integrate $p(\phi)$, $q(\phi)$ and $r(\phi)$ to $\sigma_c$ which is standard deviation of gaussian distribution denotes the amplitude of received signal and is a function. In array aspect, $\sigma_c$ is a effective variable represents the characteristic of lens, thus we call it \emph{curvature} of lens. The relationship between curvature and $\sigma_c$ is reciprocal proportion. In other words, concentration of amplitude is maximized with large curvature and minimized with small curvature.

The relationship between parameters determines characteristic of lens and curvature is experimentally given in Table \ref{table}. The relationship between parameters is experimentally confirmed by beam propagation method.
For simplicity, we define $\bf{A}\left(\phi\right)$ is $N \times N$ diagonal matrix with $\left(n,n\right)$ element follows Gaussian distribution with ${\bf{A}}\left( \phi \right)_{n,n} \sim \mathcal{N}\left( \frac{N-1}{\pi}\phi , \sigma_{c} \right)$. Then only mean changes linearly from end to end of array according to DoA, i.e., $q(\theta)=\phi(N-1)/\pi$ and $r(\phi)=\sigma_c^2$.
\begin{equation}\label{eq:a_matrix}
{\bf{A}}{\left( \phi  \right)_{n,n}} = {\sqrt {{p_{lens}}} }\frac{1}{{\sqrt {2\pi \sigma _c^2} }}{e^{ - \frac{{{{\left( {  n + \frac{{N - 1}}{\pi }\phi } \right)}^2}}}{{\sigma _c^2}}}}
\end{equation}

Here, $p_{lens}$ is a power normalizing factor which guarantees equal power gain between with and without lens. Define $p_{lens}$ in DoA average sense to be fair.
\begin{equation} \label{p_lens}
{\rm{E}}_{\phi} \left[ \sum{{\bf{A}}{\left( \phi  \right)_{n,n}}}^2 \right] =N
\end{equation}



\section{System Model} \label{sec:system model}
\begin{figure} [!t] 
\centering
\includegraphics[width=1\columnwidth] {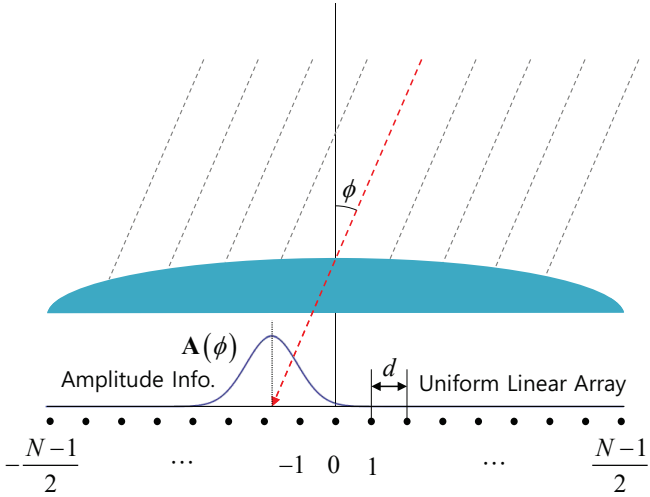}
\caption{System Model}
\label{fig:system model}
\end{figure}

Let us consider an array antenna covered with lens depicted in Fig. \ref{fig:system model}. Array antenna has odd number of elements, $N$, equispaced by $d$ with a signal coming from $\phi$ direction. The index of each element is set to be from $-{N-1}/{2}$ to ${N-1}/{2}$ to make the index of center $0$. ${\bf{x}}$ is ${N \times 1}$ column received signal vector, $p$ is signal amplitude and $b$ is phase without lens. $f\left( {{\sigma _c}} \right)$ is phase induced by characteristic of lens, which is deterministic value determined by phase transform function of the lens. ${\bf{A}}\left( \phi \right)$ is amplitude shape derived by lens. $\bf{n}$ is additive white gaussian noise (AWGN) with noise variance $\sigma_n$.

\begin{equation} \label{x}
{\bf{x}} = p{\bf{A}}\left( \phi  \right){e^{j\left\{ {b + f\left( {{\sigma _c}} \right)} \right\}}}{\bf{s}}\left( \phi  \right) + {\bf{n}}
\end{equation}

 ${\bf{s}(\phi)}$ represents the steering vector of the signal with direction $\phi$. For convenience, we set phase reference is at the center of the array by using $z = {e^{jkd\sin \phi }}$ where $k$ is time index.
\begin{equation}\label{s(phi)}
{\bf{s}}\left( \phi  \right) = \left[ 
{\begin{array}{*{7}{c}} {{z^{ - \left( {N - 1} \right)/2}}} \\ {{z^{ - \left( {N - 3} \right)/2}}} \\ \vdots \\ 1 \\ \vdots \\ {{z^{\left( {N - 3} \right)/2}}}
\\ {{z^{\left( {N - 1} \right)/2}}} \end{array}} 
\right] 
\end{equation}

\section{CRLB Analysis} \label{sec:crlb}
 In this section, we derive the bound of minimum error variance for estimating direction of arrival in terms of CRLB. The minimum error variance of any unbiased estimator is given by the CRLB in \cite{Kay:2013vs}.
 
\begin{lemma} [Cram$\acute{\rm e}$r Rao Lower Bound]
Suppose $\theta$ is an unknown deterministic parameter estimated by measurements $x$ with given probability density function $f(x|\theta)$. Then the variance of unbiased estimator $\hat{\theta}$ is bounded by the reciprocal of the Fisher information $I(\theta)$.
\begin{equation}\label{CRLB_theorem}
var(\hat{\theta}) \ge \frac{1}{I(\theta)}
\end{equation}

The Fisher information is given as follows:
\begin{equation}\label{fisher information}
I(\theta)=-{\rm{E}} \left[ \frac{\partial ^2\ln\left(f(x|\theta)\right)}{\partial \theta^2} \right]
\end{equation}
\end{lemma}

\subsection{CRLB for DoA Estimation of Linear Array}
$\bf{A} \left( \phi \right)$ becomes identity matrix without lens. Broadly it is called ULA, and CRLB of it is derived in 
\begin{equation}\label{CRLB_ULA}
CRLB_{no \ lens}\left( \phi  \right) = \frac{{6\sigma _n^2}}{{{p^2}N\left( {{N^2} - 1} \right){k^2}{d^2}{{\cos }^2}\phi }}
\end{equation}

\subsection{CRLB Analysis for DoA Estimaiton with Lens-Assisted Antenna Array}
We define three parameters into a vector, $\bf{\theta}=\left[ {\begin{array}{*{20}{c}}p&b&\phi\end{array}} \right]$, includes amplitude, phase and direction of arrival. Note that $\phi$ is a parameter estimated from the received data.
By setting ${\bf{v}} = p{\bf{A}}\left( \phi  \right){e^{j\left\{ {b + f\left( {{\sigma _c}} \right)} \right\}}}{\bf{s}}\left( \phi  \right)$, the probability density function given parameter vector $\bf{\theta}$ is represented as follows.
\begin{equation}\label{pdf_origin}
{f_{\bf{x}}}\left( {{\bf{x}}|{\bf{\theta }}} \right) = C{e^{ - {{\left( {{\bf{x}} - {\bf{v}}} \right)}^{\rm{H}}}{{\bf{R}}^{ - 1}}\left( {{\bf{x}} - {\bf{v}}} \right)}}
\end{equation}

 Where $\bf{R}=\sigma^{2}_{n} \bf{I}$ and $C$ is a normalization constant. The log-likelihood of \eqref{pdf_origin} without additional constant is
\begin{equation} \label{pdf_mod}
\begin{split}
g\left( {\bf{\theta }} \right) =  \frac{1}{{\sigma _n^2}}  & \left[  p{e^{ - j\left\{ {b + f\left( {{\sigma _c}} \right)} \right\}}}{{\bf{s}}^{\rm{H}}}\left( \phi  \right){\bf{A}}\left( \phi  \right){\bf{x}} \right . \\
& + p{e^{j\left\{ {b + f\left( {{\sigma _c}} \right)} \right\}}}{{\bf{x}}^{\rm{H}}}{\bf{s}}\left( \phi  \right){\bf{A}}\left( \phi  \right) \\
& \left .  - {p^2}{{\bf{s}}^{\rm{H}}}\left( \phi  \right){{\bf{A}}^2}\left( \phi  \right){\bf{s}}\left( \phi  \right) \right]
\end{split}
\end{equation}

Due to the fact that not all of off-diagonal terms in the Fisher information matrix are zero, we should calculate each element of fisher information matrix which is,
\begin{equation} \label{second derivative}
\begin{split}
{\rm{E}}\left[ {\frac{{{\partial ^2}g}}{{\partial {p^2}}}} \right] = & - \frac{2}{{\sigma _n^2}}{{\bf{s}}^{\rm{H}}}\left( \phi  \right){{\bf{A}}^2}\left( \phi  \right){\bf{s}}\left( \phi  \right)\  \\
{\rm{E}}\left[ {\frac{{{\partial ^2}g}}{{\partial {b^2}}}} \right] = & - \frac{{2{p^2}}}{{\sigma _n^2}}{{\bf{s}}^{\rm{H}}}\left( \phi  \right){{\bf{A}}^2}\left( \phi  \right){\bf{s}}\left( \phi  \right) \\
{\rm{E}}\left[ {\frac{{{\partial ^2}g}}{{\partial {\phi ^2}}}} \right] = & - \frac{{2{p^2}}}{{\sigma _n^2}} \frac{{4{{\left( {N - 1} \right)}^2}}}{{{\pi ^2}\sigma _c^4}}{{\bf{s}}^{\rm{H}}}\left( \phi  \right){{\bf{C}}^2}\left( \phi  \right){{\bf{A}}^2}\left( \phi  \right){\bf{s}}\left( \phi  \right)  \\
& - \frac{{2{p^2}}}{{\sigma _n^2}} {\bf{s}}_1^{\rm{H}}\left( \phi  \right){{\bf{A}}^2}\left( \phi  \right){{\bf{s}}_1}\left( \phi  \right) \\
{\rm{E}}\left[ {\frac{{{\partial ^2}g}}{{\partial p\partial b}}} \right] = & 0 \\
{\rm{E}}\left[ {\frac{{{\partial ^2}g}}{{\partial b \partial \phi}}} \right] = & \frac{{2j{p^2}}}{{\sigma _n^2}}{{\bf{s}}^{\rm{H}}}\left( \phi  \right){{\bf{A}}^2}\left( \phi  \right){{\bf{s}}_1}\left( \phi  \right) \\
{\rm{E}}\left[ {\frac{{{\partial ^2}g}}{{\partial \phi \partial p}}} \right] = & \frac{{4p\left( {N - 1} \right)}}{{\pi \sigma _n^2\sigma _c^2}}{{\bf{s}}^{\rm{H}}}\left( \phi  \right){\bf{C}}\left( \phi  \right){{\bf{A}}^2}\left( \phi  \right){\bf{s}}\left( \phi  \right) 
\end{split}
\end{equation}
where ${\bf{s}_1}$ is partial derivative of ${\bf{s}}$, ${\bf{s}_1}\left( \phi  \right) = { \partial {{\bf{s}}\left( \phi  \right)}}/{ \partial \phi}$. ${\bf{C}}{\left( \phi  \right)_{n,n}} = n + \frac{{N - 1}}{\pi }\phi$ is defined for simple expression of $ \partial {\bf{A}} / \partial \phi $. For simplicity, \eqref{Ds} is used.
\begin{equation} \label{Ds}
\begin{split}
D\left( \phi  \right) & = \frac{{{p_{lens}}}}{{2\pi \sigma _c^2}} \sum\limits_{n=-(N-1)/2}^{(N-1)/2} {{e^{ - \frac{{2{{\left( {  n + \frac{{N - 1}}{\pi }\phi } \right)}^2}}}{{\sigma _c^2}}}}} \\
{D_1}\left( \phi  \right) & = \frac{{{p_{lens}}}}{{2\pi \sigma _c^2}}\sum\limits_{n=-(N-1)/2}^{(N-1)/2} {n{e^{ - \frac{{2{{\left( { n + \frac{{N - 1}}{\pi }\phi }\right)}^2}}}{{\sigma _c^2}}}}} \\
{D_2}\left( \phi  \right) & = \frac{{{p_{lens}}}}{{2\pi \sigma _c^2}}\sum\limits_{n=-(N-1)/2}^{(N-1)/2} {{n^2}{e^{ - \frac{{2{{\left( {  n + \frac{{N - 1}}{\pi }\phi } \right)}^2}}}{{\sigma _c^2}}}}}\\
{D_{frac}}\left( \phi  \right) & = \frac{D\left( \phi  \right)}{{D\left( \phi  \right){D_2}\left( \phi  \right) - {D_1}\left( \phi  \right){D_1}\left( \phi  \right)}}
\end{split}
\end{equation}

$D_{frac}$ exists always and is finite since ${{D\left( \phi  \right){D_2}\left( \phi  \right) > {D_1}\left( \phi  \right){D_1}\left( \phi  \right)}}$. Proof is provided in Appendix. Thus, determinant of the Fisher information matrix ${\bf{J}}(\theta)$ which is nonzero always exists and can be derived based on \eqref{Ds}. 
\begin{equation} \label{determinant}
\begin{split}
\det \left\{ {{\bf{J}}\left( {\bf{\theta }} \right)} \right\} = & \frac{{8{p^4}}}{{\sigma _n^6}}D\left( \phi  \right)\left[ {{D_1}\left( \phi  \right){D_1}\left( \phi  \right) - D\left( \phi  \right){D_2}\left( \phi  \right)} \right] \\
& \left[ {\frac{{4{{\left( {N - 1} \right)}^2}}}{{{\pi ^2}\sigma _c^4}} + {k^2}{d^2}{{\cos }^2}\phi } \right]
\end{split}
\end{equation}

The CRLB for DoA estimation with lens-assisted array antenna always exists as follows.
\begin{theorem} [CRLB for DoA estimation of lens antenna]
The error variance for the DoA estimation problem of lens antenna is lower bounded by $CRLB_{lens}(\phi)$
\begin{equation} \label{CRLB_lens}
CRLB_{with \ lens}(\phi) =  \frac{D_{frac}\left( \phi  \right){\sigma _n^2}}{{ 2{p^2} \left[ {\frac{{4{{\left( {N - 1} \right)}^2}}}{{{\pi ^2}\sigma _c^4}} + {k^2}{d^2}{{\cos }^2}\phi } \right]}}
\end{equation}
\end{theorem}

By definition of $p_{lens}$ and $D\left( \phi \right)$, $D\left( \phi \right) \approx N$ is reasonable approximation since it is sum of received power. For large $\sigma_c$, i.e., when curvature of lens is small, $\frac{{4{{\left( {N - 1} \right)}^2}}}{{{\pi ^2}\sigma _c^4}} \approx 0$ and $e^{ -{ \left(  -\frac{N-1}{2} -\frac{N-1}{\pi}\phi  \right)^2}/{ \sigma_{c}^{2}} } \approx 1$. Then,
\begin{equation}
\begin{split}
{D_1}\left( \phi  \right) & \approx \sum\limits_{n=-(N-1)/2}^{(N-1)/2} {n\frac{{{p_{lens}}}}{{2\pi \sigma _c^2}}} \approx 0\\
{D_2}\left( \phi  \right) & \approx \sum\limits_{n=-(N-1)/2}^{(N-1)/2} {{n^2}\frac{{{p_{lens}}}}{{2\pi \sigma _c^2}}} \approx \sum \limits_{n=-(N-1)/2}^{(N-1)/2}{n^2} \\
& = \frac{N(N^2-1)}{12}
\end{split}
\end{equation}

Therefore \eqref{CRLB_lens} with large $\sigma_c$ can be approximated
\begin{equation} \label{approx}
CRLB_{with \ lens}\left( \phi \right) \approx  \frac{{6\sigma _n^2}}{{{p^2}N\left( {{N^2} - 1} \right){k^2}{d^2}{{\cos }^2}\phi }}
\end{equation}

 \eqref{approx} is identical to \eqref{CRLB_ULA}. It implies that DoA estimation based on lens antenna with small curvature will show similar performance without lens which is intuitively understandable.

\begin{figure} [!t] 
\centering
\includegraphics[width=1\columnwidth] {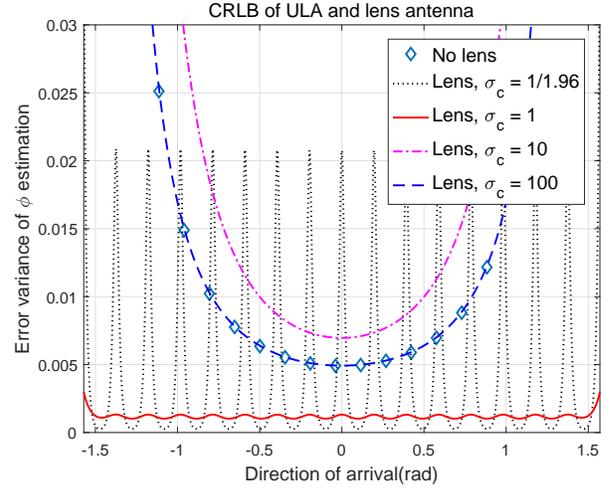}
\caption{CRLB comparison with lens antenna and ULA}
\label{fig:graph}
\end{figure}

\subsection{Graphical Description of CRLB} \label{sec:numerical results}
In this section, we compared CRLB of antenna with and without lens according to result of previous chapter. The number of antenna is 17 and $\sigma_c$ is selected from $1/1.96$ from $100$ to represent various curvature of lens.

\subsubsection{Small curvature}
 By setting $\sigma_c$ to $100$ and $10$, lens with small curvature is modeled. Since power is normalized, the performance of it is expected to be similar to that without lens, and it is shown by the simulation, which also confirms that derivation of CRLB for DoA estimation with lens is proper. When $\sigma_{c}=10$, the bound goes up which means received SNR is not properly focused to boost DoA estimation.

\subsubsection{Proper curvature}
 With proper curvature of lens, estimation error variance decreases greatly. This implies that the optimum curvature in aspect of DoA estimation can exist.

\subsubsection{Large curvature}
 The standard deviation of large curvature is set to be $1/1.96$ to depict that power from center of maximum amplitude to $d$ length for both sides of antenna is $95\%$ of total received power. Due to the system model that mean of Gaussian slides linearly according to direction, $95\%$ of total power is received by one or two elements only. Intuitively performance will be degraded more when almost whole power is received by a element. It is well shown in Fig. \ref{fig:graph}. When mean of Gaussian is nearby the antenna element, i.e., almost whole power is concentrated to a antenna, bound goes up through the case without lens, however, error bound is extremely small in case of mean positioning between two elements on the other hand.

\section{Conclusion} \label{sec:conclusion}
In this letter, we investigated a CRLB of lens embedded antenna array with proper model based on experimental results of previous research. Our results demonstrate that lens antenna has potential to show better performance in an aspect of DoA estimation than without lens.
The algorithm achieves the bound should be studied which calls for further future research.

\section{Appendix}

\begin{proof}
Here we define $f({p,m})$ for simple expression of ${{D\left( \phi  \right){D_2}\left( \phi  \right) > {D_1}\left( \phi  \right){D_1}\left( \phi  \right)}}$.
\begin{equation}
f\left( {p,m} \right) = \sum\limits_{k =  - p}^p {{k^m}{a_k}} , where' {a_k} = {e^{ - \frac{{2{{\left( {k + n} \right)}^2}}}{{\sigma _c^2}}}}
\end{equation}

Then \eqref{Ds} can be represented if $n=\phi(N-1)/\pi$. Note that ${{D\left( \phi  \right){D_2}\left( \phi  \right) > {D_1}\left( \phi  \right){D_1}\left( \phi  \right)}}$ is true when $p=1$. That is,
\begin{equation}
f\left( {1,2} \right)f\left( {1,0} \right) - {\left\{ {f\left( {1,1} \right)} \right\}^2} = 4{a_{ - 1}}{a_1} + {a_{ - 1}}{a_0} + {a_0}{a_1} > 0
\end{equation}

Suppose the inequality holds when $p=\ell$, which is $f\left( {\ell,2} \right)f\left( {\ell,0} \right) - {\left\{ {f\left( {\ell,1} \right)} \right\}^2} > 0$. Then, $p=\ell+1$ case can be shown in terms of \eqref{l+1}
\begin{equation}\label{l+1}
\begin{split}
f\left( {\ell + 1,0} \right) = &{a_{ - \left( {\ell + 1} \right)}} + f\left( {\ell,0} \right) + {a_{\left( {\ell + 1} \right)}}\\
f\left( {\ell + 1,1} \right) = & - \left( {\ell + 1} \right){a_{ - \left( {\ell + 1} \right)}} + f\left( {\ell,1} \right) + \left( {\ell + 1} \right){a_{\left( {\ell + 1} \right)}}\\
f\left( {\ell + 1,2} \right) = &{\left( {\ell + 1} \right)^2}{a_{ - \left( {\ell + 1} \right)}} + f\left( {\ell,2} \right) + {\left( {\ell + 1} \right)^2}{a_{\left( {\ell + 1} \right)}}
\end{split}
\end{equation}

The inequality when $p=\ell+1$ can be expressed,
\begin{equation}\label{p=l+1}
\begin{split}
&f\left( {\ell + 1,2} \right)f\left( {\ell + 1,0} \right) - {\left\{ {f\left( {\ell + 1,1} \right)} \right\}^2}\\
&= f\left( {\ell,2} \right)f\left( {\ell,0} \right) - f\left( {\ell,1} \right)f\left( {\ell,1} \right) + 4{\left( {\ell + 1} \right)^2}{a_{\left( {\ell + 1} \right)}}{a_{ - \left( {\ell + 1} \right)}}\\
&+ \left[ {f\left( {\ell,2} \right) + 2\left( {\ell + 1} \right)f\left( {\ell,1} \right) + {{\left( {\ell + 1} \right)}^2}f\left( {\ell,0} \right)} \right]{a_{ - \left( {\ell + 1} \right)}}\\
&+ \left[ {f\left( {\ell,2} \right) - 2\left( {\ell + 1} \right)f\left( {\ell,1} \right) + {{\left( {\ell + 1} \right)}^2}f\left( {\ell,0} \right)} \right]{a_{\left( {\ell + 1} \right)}}
\end{split}
\end{equation}

If $f\left( {\ell,1} \right) > 0$, the mean of Gaussian is negative number so that ${a_{\left( {\ell + 1} \right)}} > {a_{ - \left( {\ell + 1} \right)}}$ holds. Thus we can express $f(\ell,1)$ as $\sqrt{f(\ell,2)f(\ell,0)}-\varepsilon$ with some positive constant $\varepsilon$.
\begin{equation}\label{finaleq}
\begin{split}
&f\left( {\ell + 1,2} \right)f\left( {\ell + 1,0} \right) - {\left\{ {f\left( {\ell + 1,1} \right)} \right\}^2}\\
&= \left[ {f\left( {\ell,2} \right)f\left( {\ell,0} \right) - f\left( {\ell,1} \right)f\left( {\ell,1} \right)} \right] + 4{\left( {\ell + 1} \right)^2}{a_{\left( {\ell + 1} \right)}}{a_{ - \left( {\ell + 1} \right)}}\\
&+ {\left[ {\sqrt {f\left( {\ell,2} \right)}  + \left( {\ell + 1} \right)\sqrt {f\left( {\ell,0} \right)} } \right]^2}{a_{ - \left( {\ell + 1} \right)}}\\
&+ {\left[ {\sqrt {f\left( {\ell,2} \right)}  - \left( {\ell + 1} \right)\sqrt {f\left( {\ell,0} \right)} } \right]^2}{a_{\left( {\ell + 1} \right)}}\\
&+ 2\left( {\ell + 1} \right)\varepsilon \left\{ {{a_{\left( {\ell + 1} \right)}} - {a_{ - \left( {\ell + 1} \right)}}} \right\}
\end{split}
\end{equation}

Since \eqref{finaleq} is sum of positive terms, it is greater than zero. In case of $f\left( {\ell,1} \right) < 0$, same process with opposite sign shows inequality holds.
\end{proof}
\bibliographystyle{IEEEtran}
\bibliography{IEEEabrv,ref_crlb}

\end{document}